\documentclass[11pt]{llncs}
\usepackage{graphicx}
\usepackage{cite}
\usepackage{url}
\usepackage{moretext}

\usepackage{listings}
\mathcode`O="724F

\let\oldendproof\endproof
\def\endproof{\qed\oldendproof}

\begin{document}

\pagestyle{headings}
\mainmatter

\title{The Traveling Salesman Problem\\
for Cubic Graphs}
\titlerunning{TSP for Cubic Graphs}

\author{David Eppstein\thanks{Work supported in part by NSF grant CCR-9912338.
A preliminary version of this paper appeared at the 8th Annual Workshop on Algorithms and Data Structures, Ottawa, 2003.}}
\authorrunning{David Eppstein}
\institute{School of Information \& Computer Science\\
University of California, Irvine\\
Irvine, CA 92697-3425, USA\\
\email{eppstein@ics.uci.edu}}

\maketitle

\begin{abstract}
We show how to find a Hamiltonian cycle in a graph of degree at most three with $n$ vertices, in time
$O(2^{n/3})\approx 1.260^n$ and linear space.  Our algorithm can find the minimum weight Hamiltonian cycle (traveling salesman problem), in the same time bound.
We can also count or list all Hamiltonian cycles in a degree three graph
in time $O(2^{3n/8})\approx 1.297^n$.
We also solve the traveling salesman problem in graphs of degree at most four, by randomized and deterministic algorithms with runtime $O((27/4)^{n/3})\approx 1.890^n$ and $O((27/4+\epsilon)^{n/3})$ respectively.  Our algorithms allow the input to specify a set of forced edges which must be part of any generated cycle.
Our cycle listing algorithm shows that every degree three graph has $O(2^{3n/8})$ Hamiltonian cycles; we also exhibit a family of graphs with $2^{n/3}$ Hamiltonian cycles per graph.
\end{abstract}

\section{Introduction}

The traveling salesman problem and the closely related Hamiltonian cycle problem are two of the most fundamental of NP-complete graph problems~\cite{GarJoh-79}. However, despite much progress on exponential-time solutions to other graph problems such as chromatic number~\cite{Bys-TR-02,Epp-WADS-01,Law-IPL-76} or maximal independent sets~\cite{Bei-SODA-99,Rob-Algs-86,TarTro-SJC-77}, the only worst-case bound known for finding Hamiltonian cycles or traveling salesman tours is that for a simple dynamic program, using time and space  $O(2^n n^{O(1)})$, that finds Hamiltonian paths with specified endpoints for each induced subgraph of the input graph (D. S. Johnson, personal communication).  Hamiltonian cycle and TSP heuristics without worst case analysis have also been studied extensively~\cite{JohMcG-LSCO-97,Van-MS-98}.  Therefore, it is of interest to find special cases of the problem that, while still NP-complete, may be solved more quickly in the worst case than the general problem.

In this paper, we consider one such case: the traveling salesman problem in graphs with maximum degree three, which arises e.g. in computer graphics in the problem of stripification of triangulated surface models~\cite{ArkHelMit-VC-96,EppGop-EG-04}.  Bounded-degree maximum independent sets had previously been considered~\cite{Bei-SODA-99} but we are unaware of similar work for the traveling salesman problem in bounded degree graphs.  More generally, we consider the {\em forced traveling salesman problem} in which the input is a multigraph $G$ and set of {\em forced edges} $F$; the output is a minimum cost Hamiltonian cycle of $G$, containing all edges of $F$.  A naive branching search that repeatedly adds one edge to a growing path, choosing at each step one of two edges at the path endpoint, and backtracking when the chosen edge leads to a previous vertex, solves this problem in time $O(2^n)$ and linear space; this is already an improvement over the general graph dynamic programming algorithm.  One could also use algorithms for listing all maximal independent sets in the line graph of the input, to solve the problem in time $O(3^{n/2})\approx 1.732^n$.  We show that more sophisticated backtracking can solve the forced traveling salesman problem (and therefore also the traveling salesman and Hamiltonian cycle problems) for cubic graphs in time $O(2^{n/3})\approx 1.260^n$ and linear space.  We also provide a randomized reduction from degree four graphs to degree three graphs solving the traveling salesman problem in better time than the general case for those graphs.
We then consider the problem of listing all Hamiltonian cycles.
We show that all such cycles can be found in  time $O(2^{3n/8})\approx 1.297^n$ and linear space.
We can also count all Hamiltonian cycles in the same time bound.
Our proof implies that every degree three graph has $O(2^{3n/8})$ Hamiltonian cycles; we do not know whether this bound is tight, but we exhibit an infinite family of graphs with $2^{n/3}$ Hamiltonian cycles per graph.

\section{The Algorithm and its Correctness}

Our algorithm is based on a simple case-based backtracking technique.
Recall that $G$ is a graph with maximum degree~3, while $F$ is a set of edges that must be used in our traveling salesman tour.  For simplicity, we describe a version of the algorithm that returns only the {\em cost} of the optimal tour, or the special value {\em None} if there is no solution.  The tour itself can be reconstructed by keeping track of which branch of the backtracking process led to the returned cost; we omit the details
The steps of the algorithm are listed in Table~\ref{tbl:alg}.
Roughly, our algorithm proceeds in the following stages.  Step~\ref{step:simplify} of the algorithm reduces the size of the input without branching, after which the graph can be assumed to be cubic and triangle-free, with forced edges forming a matching.
Step~\ref{step:4cycle} tests for a case in which all unforced edges form disjoint 4-cycles;
we can then solve the problem immediately via a minimum spanning tree algorithm.
Finally (steps \ref{step:branch}--\ref{step:return}), we choose an edge to branch on, and divide the solution space into two subspaces, one in which the edge is forced to be in the solution and one in which it is excluded.  These two subproblems are solved recursively, and it is our goal to minimize the number of times this recursive branching occurs.

\begin{table}[tp]
\begin{enumerate}
\item\label{step:simplify}
Repeat the following steps until one of the steps returns or none of them applies:
\begin{enumerate}
\item\label{step:simplify-d01}
If $G$ contains a vertex with degree zero or one, return {\em None}.
\item\label{step:force-d2}
If $G$ contains a vertex with degree two, add its incident edges to $F$.
\item\label{step:simplify-ham}
If $F$ consists of a Hamiltonian cycle, return the cost of this cycle.
\item\label{step:simplify-nonham}
If $F$ contains a non-Hamiltonian cycle, return {\em None}.
\item\label{step:simplify-claw}
If $F$ contains three edges meeting at a vertex, return {\em None}.
\item\label{step:contract-d2}
If $F$ contains exactly two edges meeting at some vertex, remove from $G$ that vertex and any other edge incident to it; replace the two edges by a single forced edge connecting their other two endpoints,
having as its cost the sum of the costs of the two replaced edges' costs.
\item\label{step:simplify-parallel}
If $G$ contains two parallel edges, at least one of which is not in $F$, and $G$ has more than two vertices, then remove from $G$ whichever of the two edges is unforced and has larger cost.
\item\label{step:simplify-loop}
If $G$ contains a self-loop which is not in $F$, and $G$ has more than one vertex, remove the self-loop from $G$.
\item\label{step:contract-tri}
f $G$ contains a triangle $xyz$, then for each non-triangle edge $e$ incident to a triangle vertex, increase the cost of $e$ by the cost of the opposite triangle edge.  Also, if the triangle edge opposite $e$ belongs to $F$, add $e$ to $F$.  Remove from $G$ the three triangle edges, and contract the three triangle vertices into a single supervertex.
\item\label{step:force-quad}
If $G$ contains a cycle of four unforced edges, two opposite vertices of which are each incident to a forced edge outside the cycle, then add to $F$ all non-cycle edges that are incident to a vertex of the cycle.
\end{enumerate}
\item\label{step:4cycle}
If $G\setminus F$ forms a collection of disjoint 4-cycles, perform the following steps.
\begin{enumerate}
\item\label{step:4cycle-cheapest}
For each 4-cycle $C_i$ in $G\setminus F$, let $H_i$ consist of two opposite
edges of $C_i$, chosen so that the cost of $H_i$ is less than or equal
to the cost of $C_i\setminus H_i$.
\item\label{step:4cycle-d2ss}
Let $H=\cup_i H_i$.  Then $F\cup H$ is a degree-two spanning subgraph of $G$, but may not be connected.
\item\label{step:4cycle-patchgraph}
Form a graph $G'=(V',E')$, where the vertices of $V'$ consist of the connected components of $F\cup H$.  For each set $H_i$ that contains edges from two different components $K_j$ and $K_k$, draw an edge in $E'$ between the corresponding two vertices,
with cost equal to the difference between the costs of $C_i$ and of $H_i$.
\item\label{step:4cycle-mst}
Compute the minimum spanning tree of $(G',E')$.
\item\label{step:4cycle-return}
Return the sum of the costs of $F\cup H$ and of the minimum spanning tree.
\end{enumerate}
\item\label{step:branch}
Choose an edge $yz$ according to the following cases:
\begin{enumerate}
\item\label{step:branch-4cycle}
If $G\setminus F$ contains a 4-cycle, two vertices of which are adjacent to edges in $F$,
let $y$ be one of the other two vertices of the cycle and let $yz$ be an edge of $G\setminus F$ that
does not belong to the cycle.
\item\label{step:branch-force}
If there is no such 4-cycle, but $F$ is nonempty, let $xy$ be any edge in $F$ and $yz$ be an adjacent edge in $G\setminus F$.
\item\label{step:branch-initial}
If $F$ is empty, let $yz$ be any edge in $G$.
\end{enumerate}
\item\label{step:recurse-with}
Call the algorithm recursively on $G,F\cup\{yz\}$.
\item\label{step:recurse-without}
Call the algorithm recursively on $G\setminus\{yz\},F$.
\item\label{step:return}
Return the minimum of the set of at most two numbers returned by the two recursive calls.
\end{enumerate}
\caption{Forced traveling salesman algorithm for graph $G$ and forced edge set $F$.}
\label{tbl:alg}
\end{table}

\begin{figure}[t]
\centering
\includegraphics[width=3in]{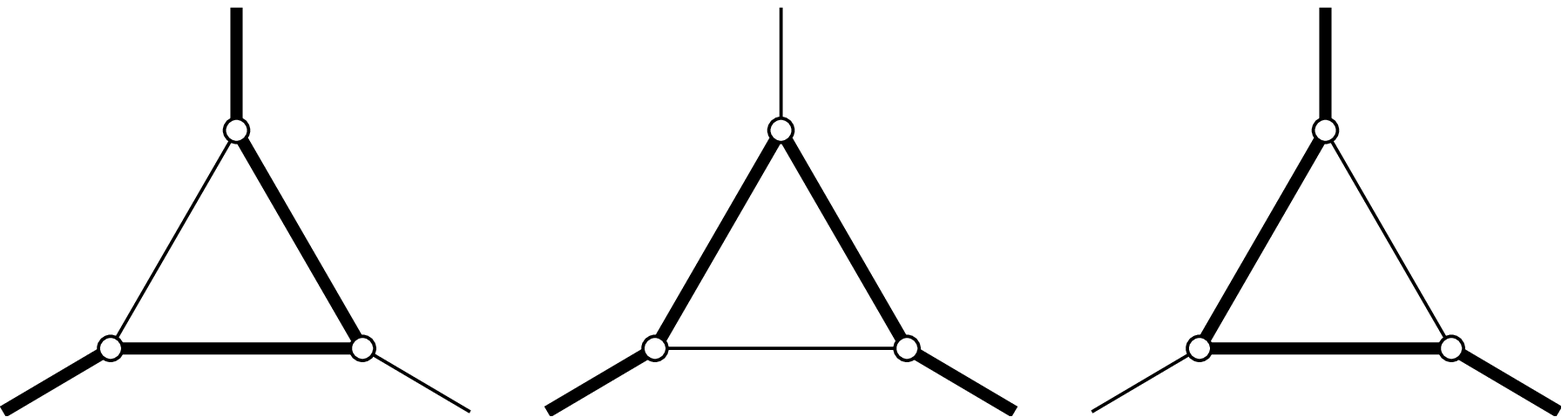}\hspace{0.5in}
\includegraphics[width=0.8in]{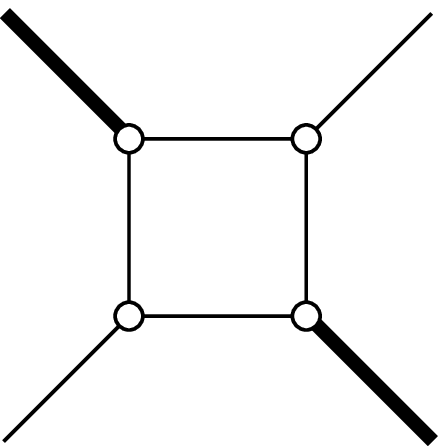}
\caption{Left: Case analysis of possible paths a Hamiltonian cycle can take through a triangle.
Edges belonging to the Hamiltonian cycle are shown as heavier than the non-cycle edges.
Right: Cycle of four unforced edges, with two forced edges adjacent to opposite cycle vertices (step~\ref{step:force-quad}).}
\label{fig:deltaham-diag4}
\end{figure}

\begin{figure}[t]
\centering
\includegraphics[width=5in]{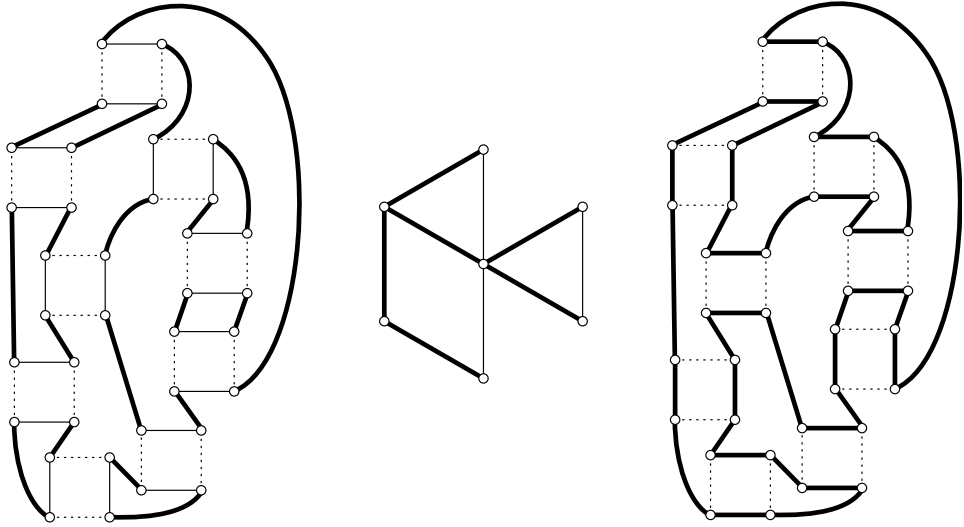}
\caption{Step~\ref{step:4cycle} of the traveling salesman algorithm.  Left: Graph with forced edges (thick lines),
such that the unforced edges form disjoint 4-cycles.  In each 4-cycle $C_i$, the pair $H_i$ of edges with lighter weight is shown as solid, and the heavier two edges are shown dashed.
Middle: Graph $G'$, the vertices of which are the connected components of solid edges in the left figure, and the edges of which connect two components that pass through the same 4-cycle.  A spanning tree of $G'$ is shown with thick lines.
Right: The tour of $G$ corresponding to the spanning tree.  The tour includes $C_i\setminus H_i$ when $C_i$ corresponds to a spanning tree edge, and includes $H_i$ otherwise.}
\label{fig:step2}
\end{figure}

All steps of the algorithm either return or reduce the input graph to one or more smaller graphs that also have maximum degree three, so the algorithm must eventually terminate.  To show correctness, each step must preserve the existence and weight of the optimal traveling salesman tour.  This is easy to verify for most cases of steps~\ref{step:simplify} and \ref{step:branch}--\ref{step:return}.
Case \ref{step:contract-tri} performs a so-called $\Delta$-$Y$ transformation on the graph; case analysis (Figure~\ref{fig:deltaham-diag4}, left)
shows that each edge of the contracted triangle participates in a Hamiltonian cycle exactly when the opposite non-triangle edge also participates.

\begin{lemma}
Let $a$, $b$, and $c$ be three edges forming a triangle in a graph $G$ in which every vertex has degree at most three, and let $d$ be a non-triangle edge incident to $b$ and $c$ (and opposite $a$).  Then every Hamiltonian cycle of $G$ that contains $a$ also contains $d$, and vice versa.
\end{lemma}

\begin{proof}
Let $T$ be the set of endpoints of the triangle.  Then a Hamiltonian cycle must enter and exit $T$ at least once.  Each entry-exit pair has two edges with one endpoint in the cycle, but there are at most three such edges overall in $G$, so the cycle can only visit $T$ once: it must enter $T$ at one of its vertices, use triangle edges to cover a second vertex, and exit $T$ from the third vertex.
The three ways the cycle might do this are shown in , and it is
easy to see from the figure that the property specified in the lemma holds for each case.
\end{proof}

This lemma justifies the correctness of adding non-triangle edges to $F$ when they are opposite forced triangle edges, and of adding the triangle edges' weights to the weights of the opposite non-triangle edges.  To justify the correctness of contracting the triangle, we use the same case analysis to prove another lemma:

\begin{lemma}
Let $a$, $b$, and $c$ be three edges forming a triangle in a graph $G$ in which every vertex has degree at most three, and let graph $G'$ be formed by removing those three edges from $G$ and replacing their endpoints by a single supervertex.  Then the Hamiltonian cycles of $G'$ are in one-to-one correspondence with the Hamiltonian cycles of $G$. If $C'$ is the set of edges in a cycle in $G'$ corresponding to a cycle $C$ in $G$, then $C'=C\setminus\{a,b,c\}$.
\end{lemma}

We now turn to step~\ref{step:force-quad}.  This step
concerns a 4-cycle in $G$, with edges in $F$ forcing the Hamiltonian cycle to enter or exit on two opposite vertices (Figure~\ref{fig:deltaham-diag4}, right).  If a Hamiltonian cycle enters and exits a cycle in $G$ only once, it does so on two adjacent vertices of the cycle, so the 4-cycle of this case is entered and exited twice by every Hamiltonian cycle, and the step's addition of edges to $F$ does not change the set of solutions of the problem.

It remains to prove correctness of step~\ref{step:4cycle} of the algorithm.

\begin{lemma}
\label{lem:step2}
Suppose that $G$, $F$ can not be reduced by step~\ref{step:simplify} of the algorithm described in Table~\ref{tbl:alg}, and that $G\setminus F$ forms a collection of disjoint 4-cycles.
Then step~\ref{step:4cycle} of the algorithm correctly solves the forced traveling salesman problem in polynomial time for $G$ and $F$.
\end{lemma}

\begin{proof}
Let $C_i$, $H_i$, $H$, and $G'$ be as defined in step~\ref{step:4cycle} of the algorithm.
Figure~\ref{fig:step2}(left) depicts $F$ as the thick edges, $C_i$ as the thin edges, 
and $H_i$ and $H$ as the thin solid edges; Figure~\ref{fig:step2}(middle) depicts the corresponding graph $G'$.

We first show that the weight of the optimal tour $T$ is at least as large as what the algorithm computes. The symmetric difference $T\oplus (F\cup H)$
contains edges only from the 4-cycles $C_i$.  Analysis similar to that for step~\ref{step:force-quad} shows that, within each 4-cycle $C_i$,
$T$ must contain either the two edges in $H_i$ or the two edges in $C_i\setminus H_i$.
Therefore, $T\oplus (F\cup H)$ forms a collection of 4-cycles which is a subset of the 4-cycles in $G\setminus F$ and which corresponds to some subgraph $S$ of $G'$.
Further, due to the way we defined the edge weights in $G'$, the difference between the weights of $T$ and of $F\cup H$ is equal to the weight of $S$.  $S$ must be a connected spanning subgraph of $G'$, for otherwise the vertices in some two components of $F\cup H$ would not be connected to each other in $T$.  Since all edge weights in $G'$ are non-negative, the weight of spanning subgraph $S$ is at least equal to that of the minimum spanning tree of~$G'$.

In the other direction, one can show by induction that, if $T'$ is any spanning tree of $G'$,
such as the one shown by the thick edges in Figure~\ref{fig:step2}(middle),
and $S'$ is the set of 4-cycles in $G$ corresponding to the edges of $T'$, then
$S'\oplus (F\cup H)$ is a Hamiltonian cycle of $G$ with weight equal to that of $F\cup H$ plus the weight of $T'$ (Figure~\ref{fig:step2}(right)).  Therefore, the weight of the optimal tour $T$  is at most equal to that of $F\cup H$ plus the weight of the minimum spanning tree of $G'$.

We have bounded the weight of the traveling salesman tour both above and below by the quantity computed by the algorithm, so the algorithm correctly solves the traveling salesman problem for this class of graphs.
\end{proof}

We summarize our results below.

\begin{theorem}
The algorithm described in Table~\ref{tbl:alg} always terminates, and returns the weight of the optimal traveling salesman tour of the input graph $G$.
\end{theorem}

\section{Implementation Details}

Define a {\em step} of the algorithm of Table~\ref{tbl:alg} to be a single execution of one of the numbered or lettered items in the algorithm description.
As described, each step involves searching for some kind of configuration in the graph, and could therefore take as much as linear time.  Although a linear factor is insignificant compared to the exponential time bound of our overall algorithm, it is nevertheless important (and will simplify our bounds) to reduce such factors to the extent possible.  As we now show, we can maintain some simple data structures that let us avoid repeatedly searching for configurations in the graph.

\begin{lemma}
\label{lem:const-per-step}
The algorithm of Table~\ref{tbl:alg} can be implemented in such a way that step~\ref{step:branch}, and each substep of step~\ref{step:simplify}, take constant time per step.
\end{lemma}

\begin{proof}
The key observation is that most of these steps require finding a connected pattern of $O(1)$ edges in the graph.  Since the graph has bounded degree, there can be at most $O(n)$ matches to any such pattern.  We can maintain the set of matches by removing a match from a set whenever one of the graph transformations changes one of its edges, and after each transformation searching within a constant radius of the changed portion of the graph for new matches to add to the set.  In this way, finding a matching pattern is a constant time operation (simply pick the first one from the set of known matches), and updating the set of matches is also constant time per operation.

The only two steps for which this technique does not work are step~\ref{step:simplify-ham} and step~\ref{step:simplify-nonham}, which each involve finding a cycle of possibly unbounded size in $G$.  However, if a long cycle of forced edges exists, step~\ref{step:simplify-claw} or step~\ref{step:contract-d2} must be applicable to the graph; repeated application of these steps will eventually either discover that the graph is non-Hamiltonian or reduce the cycle to a single self-loop.  So we can safely replace step~\ref{step:simplify-ham} and step~\ref{step:simplify-nonham} by steps that search for a one-vertex cycle in $F$, detect the applicability of the modified steps by a finite pattern matching procedure, and use the same technique for maintaining sets of matches described above to solve this pattern matching problem in constant time per step.
\end{proof}

To aid in our analysis, we restrict our algorithm so that, when it can choose among several applicable steps, it gives first priority to steps which immediately return
(that is, step~\ref{step:simplify-d01} and steps \ref{step:simplify-ham}--\ref{step:simplify-claw}, with the modifications to step~\ref{step:simplify-ham} and step~\ref{step:simplify-nonham} described in the lemma above), and second priority to step~\ref{step:contract-d2}.  The prioritization among the remaining steps is unimportant to our analysis.

\section{Analysis}

\begin{figure}[t]
\centering
\includegraphics[width=3.5in]{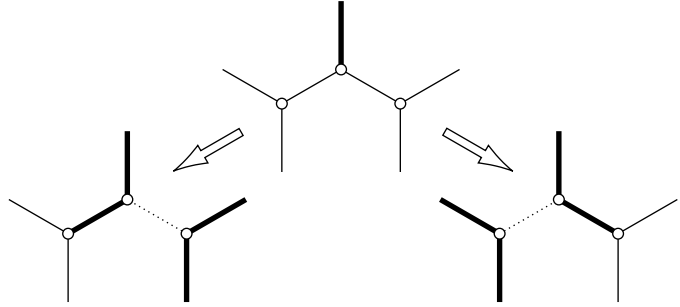}
\caption{Result of performing steps \ref{step:branch}--\ref{step:return} with no nearby forced edge: one of edges $yz$ and $yw$ becomes forced (shown as thick segments), and the removal of the other edge (shown as dotted) causes two neighboring edges to become forced.}
\label{fig:case44}
\end{figure}

By the results of the previous section, in order to compute an overall time bound for the algorithm outlined in Table~\ref{tbl:alg}, we need only estimate the number of steps it performs.  Neglecting recursive calls that immediately return, we must count the number of iterations of steps
\ref{step:force-d2}, \ref{step:contract-d2}--\ref{step:force-quad}, and \ref{step:branch}--\ref{step:return}.

\begin{lemma}\label{lem:step-1f}
If we prioritize the steps of the algorithm as described in the previous section,
the number of iterations of step~\ref{step:contract-d2} is at most $O(n)$ plus a number proportional to the number of iterations of the other steps of the algorithm.
\end{lemma}

\begin{proof}
The algorithm may perform at most $O(n)$ iterations of step~\ref{step:contract-d2} prior to executing any other step. After that point, each additional forced edge can cause at most two iterations of step~\ref{step:contract-d2}, merging that edge with previously existing forced edges on either side of it, and each step other than step~\ref{step:contract-d2} creates at most a constant number of new forced edges.
\end{proof}

The key idea of the analysis for the remaining steps is to bound the number of iterations by a recurrence involving a nonstandard measure of the size of a graph $G$:
let $s(G,F)=|V(G)|-|F|-|C|$, where $C$ denotes the set of 4-cycles of $G$ that form connected components of $G\setminus F$.  Clearly, $s\le n$, so a bound on the time complexity of our algorithm in terms of $s$ will lead to a similar bound in terms of $n$.
Equivalently, we can view our analysis as involving a three-parameter recurrence in
$n$, $|F|$, and $|C|$; in recent work~\cite{cs.DS/0304018} we showed that the asymptotic behavior of this type of multivariate recurrence can be analyzed by using weighted combinations of variables to reduce it to a univariate recurrence, similarly to our definition here of $s$ as a combination of $n$, $|F|$, and $|C|$.  Note that step~\ref{step:contract-d2} leaves $s$ unchanged and the other steps do not increase it.

\begin{lemma}\label{lem:recur}
Let a graph $G$ and nonempty forced edge set $F$ be given in which neither an immediate return nor step~\ref{step:contract-d2} can be performed, and let $s(G,F)$ be as defined above.
Then the algorithm of Table~\ref{tbl:alg}, within a constant number of steps, reduces the problem to
one of the following situations:
\begin{itemize}
\item a single subproblem $G',F'$, with $s(G',F')\le s(G,F)-1$, or
\item subproblems $G_1,F_1$ and $G_2,F_2$, with $s(G_1,F_1),s(G_2,F_2)\le s(G,F)-3$, or
\item subproblems $G_1,F_1$ and $G_2,F_2$, with $s(G_1,F_1)\le s(G,F)-2$ and $s(G_2,F_2)\le s(G,F)-5$.
\end{itemize}
\end{lemma}

\begin{proof}
If step~\ref{step:force-d2}, step~\ref{step:simplify-parallel}, step~\ref{step:simplify-loop}, or step~\ref{step:force-quad} applies, the problem is immediately reduced to a single subproblem with more forced edges, and if step~\ref{step:contract-tri} applies, the number of vertices is reduced.
Step 2 provides an immediate return from the algorithm.  So, we can restrict our attention to problems in which the algorithm is immediately forced to apply steps \ref{step:branch}--\ref{step:return}.  In such problems, the input must be a simple cubic triangle-free graph, and $F$ must form a matching in this graph, for otherwise one of the earlier steps would apply.

We now analyze cases according to the neighborhood of the edge $yz$ chosen in step~\ref{step:branch}.  To help explain the cases, we let $yw$ denote the third edge of $G$ incident to the same vertex as $xy$ and $yz$.  We also assume that no immediate return is performed within $O(1)$ steps of the initial problem, for otherwise we would again have reduced the problem to a single smaller subproblem.
\begin{itemize}
\item In the first case, corresponding to step~\ref{step:branch-4cycle} of the algorithm, $yz$ is adjacent to a 4-cycle in $G\setminus F$ which already is adjacent to two other edges of $F$.  Adding $yz$ to $F$ in the recursive call in step 4 leads to a situation in which step~\ref{step:force-quad} applies, adding the fourth adjacent edge of the cycle to $F$ and forming a 4-cycle component of $G\setminus F$.  Thus $|F|$ increases by two and $|C|$ increases by one.
In step~\ref{step:recurse-without}, $yz$ is removed from $F$, following which step~\ref{step:force-d2} adds two edges of the 4-cycle to $F$, step~\ref{step:contract-d2} contracts these two edges to a single edge, shrinking the 4-cycle to a triangle,
and step~\ref{step:contract-tri} contracts the triangle to a single vertex, so the number of vertices in the graph is decreased by three.

\item In the next case, $yz$ is chosen by step~\ref{step:branch-force} to be adjacent to forced edge $xy$, and neither $yz$ nor $yw$ is incident to a second edge in $F$.
If we add $yz$ to $F$, an application of step~\ref{step:contract-d2} removes $yw$, and another application of step~\ref{step:force-d2} adds the two edges adjoining $yw$ to $F$, so the number of forced edges is increased by three.  The subproblem in which we remove $yz$ from $F$ is symmetric.  This case and its two subproblems are shown in
Figure~\ref{fig:case44}.

\item If step~\ref{step:branch-force} chooses edge $yz$, and $z$ or $w$ is incident to a forced edge, then with $y$ it forms part of a chain of two or more vertices, each incident to exactly two unforced edges that connect vertices in the chain.
This chain may terminate at vertices with three adjacent unforced edges
(Figure~\ref{fig:forcedchain}, left).  If it does, a similar analysis to the previous case shows that
adding $yz$ to $F$ or removing it from $G$ causes alternating members of the chain to be added to $F$ or removed from $G$, so that no chain edge is left unforced.  In addition, when
an edge at the end of the chain is removed from $G$, two adjacent unforced edges are added to $F$, so these chains generally lead to a greater reduction in size than the previous case.
The smallest reduction happens when the chain consists of exactly two vertices adjacent to forced edges.  In this case, one of the two subproblems is formed by adding two new forced edges at the ends of the chain, and removing one edge interior to the chain; it has $s(G_1,F_1)=s(G,F)-2$.
The other subproblem is formed by removing the two edges at the ends of the chain, and adding to $F$ the edge in the middle of the chain and the other unforced edges adjacent to the ends of the chain.  None of these other edges can coincide with each other without creating a 4-cycle that would have been treated in the first case of our analysis, so in this case there are five new forced edges and
$s(G_2,F_2)=s(G,F)-5$.

\begin{figure}[t]
\centering
\includegraphics[width=3.5in]{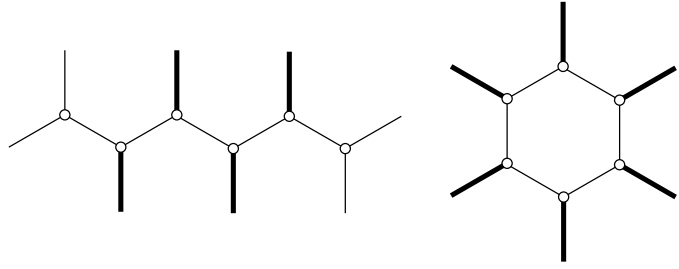}
\caption{Chains of two or more vertices each having two adjacent unforced edges.
Left: chain terminated by vertices with three unforced edges.
Right: cycle of six or more vertices with two unforced edges.}
\label{fig:forcedchain}
\end{figure}

\item In the remaining case, step~\ref{step:branch-force} chooses an edge belonging to a cycle of unforced edges, each vertex of which is also incident to a forced edge (Figure~\ref{fig:forcedchain}, right).
In this case, adding or removing one of the cycle edges causes a chain reaction which alternately adds and removes all cycle edges.  This case only arises when the cycle length is five or more, and if it is exactly five then an inconsistency quickly arises causing both recursive calls to return within a constant number of steps.  When the cycle length is six or more,
both resulting subproblems end up with
at least three more forced edges.
\end{itemize}
Note that the analysis need not consider choices made by step~\ref{step:branch-initial} of the algorithm, as $F$ is assumed nonempty; step~\ref{step:branch-initial} can occur only once and does not contribute to the asymptotic complexity of the algorithm.
In all cases, the graph is reduced to subproblems that have sizes bounded as stated in the lemma.
\end{proof}

\begin{theorem}
The algorithm of Table~\ref{tbl:alg} solves the forced traveling salesman problem on graphs of degree three in time $O(2^{n/3})$.
\end{theorem}

\begin{proof}
The algorithm's correctness has already been discussed.
By Lemmas \ref{lem:step2}, \ref{lem:const-per-step}, \ref{lem:step-1f}, and~\ref{lem:recur},
the time for the algorithm can be bounded within a constant factor by the solution to the recurrence
$$T(s)\le 1+\max\{s^{O(1)},T(s-1), 2T(s-3), T(s-2)+T(s-5)\}.$$
Standard techniques for linear recurrences give the solution as
$T(s)=O(2^{s/3})$.
In any $n$-vertex cubic graph, $s$ is at most $n$, so expressed in terms of $n$
this gives a bound of $O(2^{n/3})$ on the running time of the algorithm.
\end{proof}

\section{Degree Four}

\begin{figure}[t]
\centering
\includegraphics[width=2.5in]{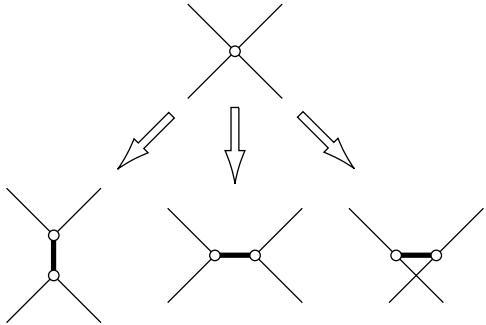}
\caption{Reducing degree four vertices to degree three vertices, by randomly splitting vertices and connecting the two sides by a forced edge.}
\label{fig:deg4split}
\end{figure}

It is natural to ask to what extent our algorithm can be generalized to higher vertex degrees.
We provide a first step in this direction, by describing a randomized (Monte Carlo) algorithm:
that is, an algorithm that may produce incorrect results with bounded probability.
To describe the algorithm, let $f$ denote the number of degree four vertices in the given graph.
The algorithm consists of $(3/2)^f$ repetitions of the following:
for each degree four vertex, choose randomly among the three possible partitions of its incoming edges into two sets of two edges; split the vertex into two vertices, with the edges assigned to one or the other vertex according to the partition, and connect the two vertices by a new forced edge
(Figure~\ref{fig:deg4split}).  Once all vertices are split, the graph has maximum degree~3 and we can apply our previous forced TSP algorithm.

It is not hard to see that each such split preserves the traveling salesman tour only when the two tour edges do not belong to the same set of the partition, which happens with probability $2/3$; therefore, each repetition of the algorithm has probability $(2/3)^f$ of finding the correct TSP solution.
Since there are $(3/2)^f$ repetitions, there is a bounded probability that the overall algorithm finds the correct solution.  Each split leaves unchanged the parameter $s$ used in our analysis of the algorithm for cubic graphs, so the time for the algorithm
is $O((3/2)^f 2^{n/3})=O((27/4)^{n/3})$.  By increasing the number of repetitions the failure probability can be made exponentially small with only a polynomial increase in runtime.

A hitting set technique due to Beigel (personal communication, 1995) allows this algorithm to be derandomized.
We group the degree four vertices arbitrarily into groups of $k$ vertices per group.
Within each group, a single choice among the three possible expansions of each degree-four vertex can be described by a $k$-digit ternary word $w\in\{0,1,2\}^k$.  If $x=x_0x_1\ldots x_{k-1}$ is a ternary word, let $D(x)=\{w_0w_1\ldots w_{k-1}\in\{0,1,2\}^k\mid \hbox{for all $i$,} w_i\neq x_i\}$.
As two of the three expansions of each vertex preserve the correct traveling salesman solution,
the set of expansions of the group that preserve the TSP solution is a set of the form $D(x)$ for some unknown $x$.  We can find one such expansion if we test all words in a {\em hitting set}: that is, a set $S$ of ternary words such that $S\cap D(x)\ne\emptyset$ for all~$x$.  In order to find a fast deterministic algorithm, we seek a hitting set that is as small as possible.

As each set $D(x)$ has the same cardinality $2^k$, the hitting set problem for the sets $D(x)$ has a simple fractional solution: assign weight $2^{-k}$ to each of the $3^k$ possible ternary words.
The total weight of this solution is therefore $(3/2)^k$.  By standard results relating integer to fractional solutions of the hitting set (or equivalently set cover) problem, this implies that there is a hitting set of cardinality at most $(3/2)^k\ln{\cal D}=O(k (3/2)^k)$, where ${\cal D}=3^k$ is the cardinality of the family of sets $D(x)$.

We use this hitting set $H$ as part of a deterministic search algorithm
that tests each choice of a member of $H$ for each group of $k$ vertices.
For each of these $|H|^{n/k}$ choices, we expand the degree four vertices in each group
as specified by the choice for that group, and then apply our degree-three TSP algorithm on the expanded graph.  At least one choice hits the set $D(x)$ in each group of expansions preserving the TSP, so the best TSP solution among the $|H|^{n/k}$ expansions must equal the TSP of the original graph.  By choosing a suitably large constant $k$, we can achieve time $O((27/4+\epsilon)^{n/3})$
for any constant $\epsilon>0$.

We omit further details as this result seems unlikely to be optimal.

\section{Listing All Hamiltonian Cycles}

Suppose we want not just a single best Hamiltonian cycle (the Traveling Salesman Problem) but rather a list of all such cycles.
As we show in Table~\ref{tbl:count}, most of the steps of our traveling salesman algorithm
can be generalized in a straightforward way to this cycle listing problem.
However, we do not know of a cycle listing analogue to the minimum spanning tree algorithm
described in step 2 of Table~\ref{tbl:alg}, and proven correct in Lemma~\ref{lem:step2}
for graphs in which the unforced edges form disjoint 4-cycles.
It is tempting to try listing all spanning trees instead of computing minimum spanning trees, however not every Hamiltonian cycle of the input graph $G$ corresponds to a spanning tree of the derived graph $G'$ used in that step.
Omitting the steps related to these 4-cycles gives the simplified algorithm shown in Table~\ref{tbl:count}.
We analyze this algorithm in a similar way to the previous one; however
in this case we use as the parameter of our analysis the number of unforced edges $U(G)$ in the graph~$G$.  Like $s(G)$, $U$ does not increase at any step of the algorithm; we now show that it decreases by sufficiently large amounts at certain key steps.

\begin{table}[t]
\begin{enumerate}
\item Repeat the following steps until one of the steps returns or none of them applies:
\begin{enumerate}
\item If $G$ contains a vertex with degree zero or one, or
if $F$ contains a non-Hamiltonian cycle or three edges meeting at a vertex, backtrack.
\item If $F$ consists of a Hamiltonian cycle, output the cycle formed by all edges of the
original input graph that have been added to $F$ and backtrack.
\item\label{step:listing-d2force}
If $G$ contains a vertex with degree two, add its incident edges to $F$.
\item\label{step:listing-triforce}
If $G$ contains a triangle $xyz$, and the non-triangle edge incident to $x$ belongs to $F$,
add edge $yz$ to $F$.
\item\label{step:listing-contract}
If $F$ contains exactly two edges meeting at some vertex, remove from $G$ that vertex and any other edge incident to it, and replace the two edges by a single edge connecting their other two endpoints.  If this contraction
would lead to two parallel edges in $G$, remove the other edge from $G$.
\end{enumerate}
\item\label{step:listing-branch}
 If $F$ is nonempty, let $xy$ be any edge in $F$ and $yz$ be an adjacent edge in $G\setminus F$.\\
Otherwise, if $F$ is empty, let $yz$ be any edge in $G$.  Call the algorithm
recursively on the two graphs $G,F\cup\{yz\}$ and  $G\setminus\{yz\},F$.
\end{enumerate}
\caption{Forced Hamiltonian cycle listing algorithm for graph $G$, forced edges $F$.}
\label{tbl:count}
\end{table}

\begin{lemma}
If $G$ initially has no parallel edges or self-loops, none are introduced by the steps of this algorithm.
\end{lemma}

\begin{proof}
The only step that adds a new edge to the graph is step~\ref{step:listing-contract}, which replaces a two-edge path with a single edge.  If the graph has no parallel edges before this step, the step cannot create a self-loop.  It can create a multiple adjacency if the two path edges belong to a triangle, but as part of the step we immediately detect and eliminate this adjacency.
\end{proof}

\begin{lemma}\label{lem:ufrecur}
Let a graph $G$ be given in which neither a backtrack nor step~\ref{step:listing-contract} can be performed, let $F$ be nonempty, and let $U(G)$ denote the number of unforced edges in $G$.
Then the algorithm of Table~\ref{tbl:count}, within a constant number of steps, reduces the problem to one of the following situations:
\begin{itemize}
\item a single subproblem $G'$, with $U(G')\le U(G)-1$, or
\item two subproblems $G_1$ and $G_2$, with $U(G_1),U(G_2)\le U(G)-4$, or
\item two subproblems $G_1$ and $G_2$, with $U(G_1)\le U(G)-3$ and $U(G_2)\le U(G)-6$.
\end{itemize}
\end{lemma}

\begin{proof}
If step~\ref{step:listing-d2force} or step~\ref{step:listing-triforce} applies, the problem is immediately reduced to a single subproblem with fewer unforced edges.  So, we can restrict our attention to problems in which the algorithm is immediately forced to apply step~\ref{step:listing-branch}.  In such problems, the input must be a simple cubic triangle-free graph, and $F$ must form a matching in this graph, for otherwise one of the earlier steps would apply.

We now perform a case analysis according to the local neighborhood of the edge $yz$ chosen in step~\ref{step:listing-triforce}.  As in Lemma~\ref{lem:recur},  let $yw$ denote the third edge of $G$ incident to the same vertex as $xy$ and $yz$.
\begin{itemize}
\item In the first case, neither $yz$ nor $yw$ is incident to a second edge in $F$.
In the subproblem in which we add $yz$ to $F$, an application of step~\ref{step:listing-contract} removes $yw$, and another application of step~\ref{step:listing-d2force} adds the two edges adjoining $yw$ to $F$.  Thus, in this case, the number of unforced edges is reduced by four.  The subproblem in which we remove $yz$ from $F$ is symmetric.  This case and the two subproblems it produces are shown in
Figure~\ref{fig:case44}.

\item If $z$ or $w$ is incident to a forced edge, then with $y$ it forms part of a chain of two or more vertices, each of which is incident to exactly two unforced edges that connect vertices in the chain.
This chain may terminate at vertices with three adjacent unforced edges
(Figure~\ref{fig:forcedchain}, left).  If it does, a similar analysis to the previous case shows that
adding $yz$ to $F$ or removing it from $G$ causes alternating members of the chain to be added to $F$ or removed from $G$ as well, so that no chain edge is left unforced.  In addition, when
an edge at the end of the chain is removed from $G$, the two adjacent unforced edges are added to $F$.
The smallest reduction in unforced edges happens when the chain consists of exactly two vertices adjacent to forced edges.  In this case, one of the two subproblems is formed by adding two new forced edges at the ends of the chain, and removing one edge interior to the chain; it thus has $U(G_1)=U(G)-3$.
The other subproblem is formed by removing the two edges at the ends of the chain, and adding to $F$ the edge in the middle of the chain and the other unforced edges adjacent to the ends of the chain.  Thus it would seem that this subproblem has $U(G_2)=U(G)-7$, however it is possible
for one unforced edge to be adjacent to both ends of the chain, in which case we only get
$U(G_2)=U(G)-6$.

\item In the remaining case, we have a cycle of four or more unforced edges, each vertex of which is also incident to a forced edge (Figure~\ref{fig:forcedchain}, right).
In this case, adding or removing one of the cycle edges causes a chain reaction which alternately adds and removes all cycle edges, so both resulting subproblems end up with
at least four fewer unforced edges.
\end{itemize}
Thus, in all cases, the graph is reduced to two subproblems that have numbers of unforced edges bounded as in the statement of the lemma.
\end{proof}

\begin{theorem}\label{thm:list-cycles}
For any simple graph $G$ with maximum degree 3, set $F$ of forced edges in $G$, and assignment of weights to the edges of $G$ from a commutative semiring, we can list all Hamiltonian cycles in $G$ in $O(2^{3n/8})$ and linear space.
\end{theorem}

\begin{proof}
By the previous lemma, the number of calls in
the algorithm can be bounded within a constant factor by the solution to the recurrence
$$T(u)\le 1+\max\{T(u-1), 2T(u-4), T(u-3)+T(u-6)\}.$$
Standard techniques for linear recurrences give the solution as
$T(u)=O(2^{u/4})$.
In any $n$-vertex cubic graph, $u$ is at most $3n/2$, so expressed in terms of $n$
this gives a bound of $O(2^{3n/8})$ on the number of operations.
As in the previous algorithm, by the appropriate use of simple data structures we can
implement each step of the algorithm in constant time per step.
\end{proof}

\begin{figure}[t]
\centering
\includegraphics[width=4in]{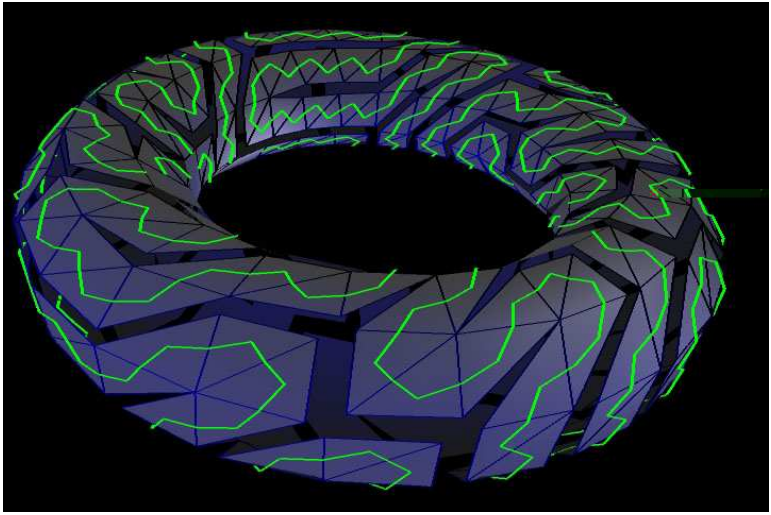}
\caption{Hamiltonian cycle of dual torus mesh, found by our implementation (courtesy M. Gopi).}
\label{fig:torus4}
\end{figure}

In order to achieve this time bound, we must output each Hamiltonian cycle using an implicit representation that changes by a constant amount in each step.  Explicitly listing the vertices in order in each Hamiltonian cycle would take $O(2^{3n/8}+Kn)$
where $K$ is the number of Hamiltonian cycles produced by the algorithm.
Appendix~A presents an implementation of this algorithm in the Python language, using such an implicit representation.
This implementation was able to find a Hamiltonian cycle of a 400-vertex 3-regular graph,
dual to a triangulated torus model (Figure~\ref{fig:torus4}) in approximately two seconds on an 800 MHz PowerPC G4 computer.  However, on a slightly larger 480-vertex graph dual to a triangulated sphere model, we aborted the computation after 11 hours with no result.

\begin{corollary}
We can count the number of Hamiltonian cycles in any cubic graph in time $O(2^{3n/8})$ and linear space.
\end{corollary}

\begin{proof}
We simply maintain a counter of the number of cycles seen so far, and increment it each time we output another cycle.  The average number of bits changed in the counter per step is $O(1)$,
so the total additional time to maintain the counter is $O(2^{3n/8})$.
\end{proof}

A preliminary version of this paper used a similar recursion for counting Hamiltonian cycles, but returned the counts from each recursive subproblem, incurring an additional polynomial factor overhead for the arithmetic involved.

\section{Graphs with Many Hamiltonian Cycles}

\begin{figure}[t]
\centering
\includegraphics[width=3.5in]{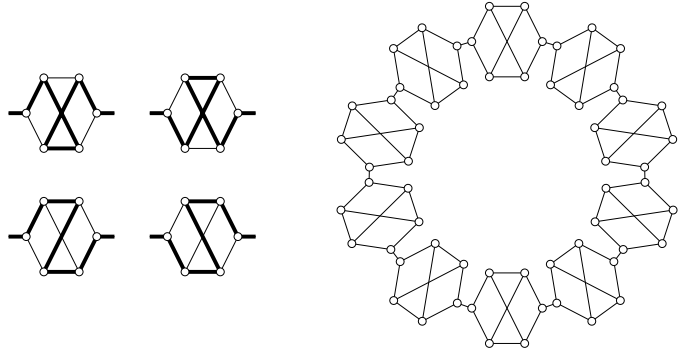}
\caption{Cubic graph with $2^{n/3}$ Hamiltonian cycles.  Left: four paths through a six-vertex gadget; right: $n/6$ gadgets connected into a cycle.}
\label{fig:manyham}
\end{figure}

The following result follows immediately from Theorem~\ref{thm:list-cycles}:

\begin{corollary}
Every simple $n$-vertex graph with maximum degree three has at most $2^{3n/8}$ Hamiltonian cycles.
\end{corollary}

For cubic multigraphs, one can show a $2^{n/2}$ bound on the number of Hamiltonian cycles by choosing one distinguished cycle, and corresponding the other cycles to subsets of the remaining $n/2$ edges.
This bound is tight as can be seen by the example of an $n$-gon with alternating single and double bonds.
We do not know whether our $2^{3n/8}$ bound is tight, but we exhibit in Figure~\ref{fig:manyham} the construction for an infinite family of graphs with $2^{n/3}$ Hamiltonian cycles per graph. 
Each graph in the family is formed by connecting $n/6$ $6$-vertex gadgets into a cycle,
where each gadget is formed by removing an edge from the graph $K_{3,3}$.
A Hamiltonian cycle of a graph formed in this way must pass once through each gadget,
and there are four possible paths through each gadget, so the total number of Hamiltonian cycles is $2^{n/3}$.  We note that the $n/6$ edges connecting pairs of gadgets in this graph are all forced to occur in any Hamiltonian cycle, so in terms of the number $u$ of unforced edges the number of Hamiltonian cycles is $4^{u/8}=2^{u/4}$, matching the worst-case bound in terms of $u$ for our cycle-listing algorithm.

We used our cycle listing algorithm to search, unsuccessfully, for a better gadget among all possible cubic graphs on 20 or fewer vertices, using tables of these graphs provided online by Gordon Royle~\cite{McKRoy-AC-86}.  Our experiments support the conjecture that, in any $n$-vertex cubic graph, each edge participates in at most $2^{\lfloor n/3\rfloor}$ Hamiltonian cycles.  This bound is achieved by the graphs of Figure~\ref{fig:manyham} as well as by similar graphs that include one or two four-vertex gadgets formed by removing an edge from $K_4$, so if true this bound would be tight.  The truth of this conjecture would imply that any cubic graph has $O(2^{n/3})$ Hamiltonian cycles.

{\raggedright
\bibliographystyle{abuser}
\bibliography{cubictsp}}

\section*{Appendix A\quad Implementation of Cycle Listing Algorithm}

We present here an implementation of the $O(2^{3n/8})$ algorithm for listing all Hamiltonian cycles in a degree three graph,  in the Python programming language~\cite{Python}.  The \lstinline@yield@ keyword triggers Python's {\em simple generator protocol}~\cite{PEP-255}, which creates an iterator object suitable for use in \lstinline@for@-loops and similar contexts and returns it from each call to \lstinline@HamiltonianCycles@.  A more elaborate version of the implementation, which backtracks when it discovers that the current graph is not biconnected or its unforced edges have no perfect matching, and includes code for testing the algorithm on several simple families of 3-regular graphs, is available for download
at \url{http://www.ics.uci.edu/~eppstein/PADS/CubicHam.py}.

\lstset{basicstyle=\ttfamily\small}
\begin{lstlisting}
"""CubicHam.py

Generate all Hamiltonian cycles in graphs of maximum degree three.
D. Eppstein, April 2004.
"""

def HamiltonianCycles(G):
    """
    Generate a sequence of all Hamiltonian cycles in graph G.
    G should be represented in such a way that "for v in G" loops through
    the vertices, and "G[v]" produces a collection of neighbors of v; for
    instance, G may be a dictionary mapping vertices to lists of neighbors.
    Each cycle is returned as a graph in a similar representation, and
    should not be modified by the caller.
    """

    # Make a copy of G so we can destructively modify it
    # G[v][w] is true iff v-w is an original edge of G
    # (rather than an edge created by a contraction of G).
    copy = {}
    for v in G:
        if len(G[v]) < 2:
            return      # Isolated or degree one vertex, no cycles exist
        copy[v] = dict([(w,True) for w in G[v]])
    G = copy
    
    # Subgraph of forced edges in the input
    forced_in_input = dict([(v,{}) for v in G])
        
    # Subgraph of forced edges in current G
    forced_in_current = dict([(v,{}) for v in G])
        
    # List of vertices with degree two
    degree_two = [v for v in G if len(G[v]) == 2]
            
    # Collection of vertices with forced edges
    forced_vertices = {}

    # The overall backtracking algorithm is implemented by means of
    # a stack of actions.  At each step we pop the most recent action
    # off the stack and call it.  Each stacked function should return None or False
    # normally, or True to signal that we have found a Hamiltonian cycle.
    # Whenever we modify the graph, we push an action undoing that modification.
    # Below are definitions of actions and action-related functions.

    def remove(v,w):
        """Remove edge v,w from edges of G."""
        was_original = G[v][w]
        del G[v][w],G[w][v]
        was_forced = w in forced_in_current[v]
        if was_forced:
            del forced_in_current[v][w],forced_in_current[w][v]
        
        def unremove():
            G[v][w] = G[w][v] = was_original
            if was_forced:
                forced_in_current[v][w] = forced_in_current[w][v] = True
        
        actions.append(unremove)

    def now_degree_two(v):
        """Discover that changing G has caused v's degree to become two."""
        degree_two.append(v)
        def not_degree_two():
            top = degree_two.pop()
        actions.append(not_degree_two)

    def safely_remove(v,w):
        """
        Remove edge v,w and update degree two data structures.
        Returns True if successful, False if found a contradiction.
        """
        if w in forced_in_current[v] or len(G[v]) < 3 or len(G[w]) < 3:
            return False
        remove(v,w)
        now_degree_two(v)
        now_degree_two(w)
        return True

    def remove_third_leg(v):
        """
        Check if v has two forced edges and if so remove unforced one.
        Returns True if successful, False if found a contradiction.
        """
        if len(G[v]) != 3 or len(forced_in_current[v]) != 2:
            return True
        w = [x for x in G[v] if x not in forced_in_current[v]][0]
        if len(G[w]) <= 2:
            return False
        return safely_remove(v,w)

    def force(v,w):
        """
        Add edge v,w to forced edges.
        Returns True if successful, False if found a contradiction.
        """
        if w in forced_in_current[v]:
            return True     # already forced
        if len(forced_in_current[v]) > 2 or len(forced_in_current[w]) > 2:
            return False    # three incident forced => no cycle exists
        forced_in_current[v][w] = forced_in_current[w][v] = True
        not_previously_forced = [x for x in (v,w) if x not in forced_vertices]
        for x in not_previously_forced:
            forced_vertices[x] = True
        was_original = G[v][w]
        if was_original:
            forced_in_input[v][w] = forced_in_input[w][v] = True
        
        def unforce():
            """Undo call to force."""
            for x in not_previously_forced:
                del forced_vertices[x]
            del forced_in_current[v][w],forced_in_current[w][v]
            if was_original:
                del forced_in_input[v][w],forced_in_input[w][v]

        actions.append(unforce)
        return remove_third_leg(v) and remove_third_leg(w) and \
            force_into_triangle(v,w) and force_into_triangle(w,v)
            
    def force_into_triangle(v,w):
        """
        After v,w has been added to forced edges, check if w
        belongs to a triangle, and if so force the opposite edge.
        Returns True if successful, False if found a contradiction.
        """
        if len(G[w]) != 3:
            return True
        x,y = [z for z in G[w] if z != v]
        if y not in G[x]:
            return True
        return force(x,y)
    
    def contract(v):
        """
        Contract out degree two vertex.
        Returns True if cycle should be reported, False or None otherwise.
        Appends recursive search of contracted graph to action stack.
        """
        u,w = G[v]
        if w in G[u]:           # about to create parallel edge?
            if len(G) == 3:     # graph is a triangle?
                return force(u,v) and force(v,w) and force(u,w)
            if not safely_remove(u,w):
                return None     # unable to remove uw, no cycles exist
        
        if not force(u,v) or not force(v,w):
            return None     # forcing the edges led to a contradiction
        remove(u,v)
        remove(v,w)
        G[u][w] = G[w][u] = False
        forced_in_current[u][w] = forced_in_current[w][u] = True
        del G[v],forced_vertices[v]
        
        def uncontract():
            del G[u][w],G[w][u]
            del forced_in_current[u][w],forced_in_current[w][u]
            forced_vertices[v] = True
            G[v] = {}

        actions.append(uncontract)
        actions.append(main)    # search contracted graph recursively

    def handle_degree_two():
        """
        Handle case that the graph has a degree two vertex.
        Returns True if cycle should be reported, False or None otherwise.
        Appends recursive search of contracted graph to action stack.
        """
        v = degree_two.pop()
        def unpop():
            degree_two.append(v)
        actions.append(unpop)
        return contract(v)

    def main():
        """
        Main event dispatcher.
        Returns True if cycle should be reported, False or None otherwise.
        Appends recursive search of contracted graph to action stack.
        """
        if degree_two:
            return handle_degree_two()
            
        # Here with a degree three graph in which the forced edges
        # form a matching.  Pick an unforced edge adjacent to a
        # forced one, if possible, else pick any unforced edge,
        # and branch on the chosen edge.
        if forced_vertices:
            v = iter(forced_vertices).next()
        else:
            v = iter(G).next()
        w = [x for x in G[v] if x not in forced_in_current[v]][0]
        
        def continuation():
            """Here after searching first recursive subgraph."""
            if force(v,w):
                actions.append(main)

        actions.append(continuation)
        if safely_remove(v,w):
            actions.append(main)

    # The main backtracking loop
    actions = [main]
    while actions:
        if actions.pop()():
            yield forced_in_input
\end{lstlisting}

 \end{document}